\documentclass{PoS}

\newcommand{\preprintline}{\newline
\vskip -4.6cm
\rightline{\parbox{4cm}{\large\rm{ITEP-LAT/2008-24}}}
\vspace{2.8cm}}

\title{
\vskip 10mm
Abelian monopoles and center vortices in Yang-Mills plasma\thanks{
This work was supported by
Grants-in-Aid for Scientific Research from "The Ministry
of Education, Culture, Sports, Science and Technology of Japan" Nos. 17340080 and 20340055,
JSPS Invitation Fellowship for Research in Japan, No. S-08035,
by the STINT Institutional grant IG2004-2 025, and by the Federal Program of the
Russian Ministry of Industry, Science and Technology No. 40.052.1.1.1112.
The numerical simulations were performed using a SX-8 supercomputer at
RCNP at Osaka University, and a SR11000 machine at Hiroshima University.}
\preprintline
}

\ShortTitle{Abelian monopoles and center vortices in Yang-Mills plasma}

\author{\speaker{M.~N.~Chernodub}\\
        Laboratoire de Mathematiques et Physique Theorique
        CNRS UMR 6083, F\'ed\'eration Denis Poisson, Universit\'e de Tours,
        Parc de Grandmont, F37200, Tours, France \\
        Institute of Theoretical and Experimental Physics ITEP, 117259
        Moscow, Russia\\
        E-mail: \email{Maxim.Chernodub@lmpt.univ-tours.fr}}
\author{Atsushi Nakamura\\
        Research Institute for Information Science and Education,
        Hiroshima University, Higashi-Hiroshima, 739-8527, Japan\\
        E-mail: \email{nakamura@riise.hiroshima-u.ac.jp}}
\author{V.~I.~Zakharov\\
        Institute of Theoretical and Experimental Physics ITEP, 117259
        Moscow, Russia\\
        Max-Planck Institut f\"ur Physik, F\"ohringer Ring 6, 80805,
        M\"unchen, Germany\\
        E-mail: \email{xxz@mppmu.mpg.de}}

\abstract{
Condensation of the Abelian monopoles and the center vortices leads to
confinement of color in low temperature phase of Yang-Mills theory.
We stress that these topological magnetic degrees of freedom are also
very important in the deconfinement regime:  at the point of the
deconfinement phase transition both the monopoles and the vortices are
released into the thermal vacuum contributing, in particular, to the
equation of state and, definitely, to transport properties of the hot gluonic
medium. Thus, we argue that a novel, magnetic component plays a crucial role.
On the other hand, it was demonstrated that an effective three-dimensional
description can be brought, beginning with high temperatures, down to
the critical temperature by postulating existence of a system of 3d Higgs
fields. We propose to identify the 3d color-singlet Higgs field with
the 3d projection of the 4d magnetic vortices. Such identification fits well
the 3d properties of the theory and contributes to interpretation
of the magnetic component of the Yang-Mills plasma.
}

\FullConference{8th Conference Quark Confinement and the Hadron Spectrum \\
         September 1-6 2008\\
         Mainz, Germany}

\newcommand{\Tr}{{\mathrm{Tr}}\,}

\newcommand{\beqn}{\begin{eqnarray}}
\newcommand{\eeqn}{\end{eqnarray}}
\newcommand{\eq}[1]{(\ref{#1})}

\newcommand{\Z}{{\mathbb Z}}

\begin{document}

\section{Introduction}

Both experimental observations at RHIC and numerical simulations of Yang-Mills theories
indicate that the Yang--Mills plasma possesses quite unusual properties, for a review see,
e.g., \cite{review}. At temperatures just above the critical temperature $T_c$, the
transport properties of the plasma correspond to an (perfect) fluid rather than to a weakly
interacting gas. A reason for the liquid nature of the gluon plasma lies in a strong
collective interaction of the gluons. A self-consistent theoretical explanation of this
phenomena is still lacking and the topic attracts great interest nowadays.

Below we further discuss the picture~\cite{chernodub,ref:review.monopoles.vortices,shuryak,nakamura,sasha,massimo}
according to which the unusual thermodynamical and transport properties of the plasma appear as due to presence
of  monopolelike and vortexlike topological defects in the gluon plasma. A brief review of the magnetic component 
of the plasma can be found in Ref.~\cite{ref:review.monopoles.vortices}.

There are two constituents of the magnetic component~\cite{chernodub}:
a particlelike magnetic monopole and a stringlike magnetic vortex. These constituents appear as
singular magnetic defects in the gluon fields. The magnetic monopoles are related to the color confinement
via the so called dual superconductor mechanism~\cite{polikarpov}. In the vortex picture the quark confinement emerges as
a result of a percolation of the vortices~\cite{greensite}.

\begin{figure}[!htb]
\begin{center}
\includegraphics[angle=-0,scale=0.75,clip=true]{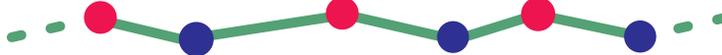}
\end{center}
\caption{A part of a monopole-vortex chain in $SU(2)$ gauge theory.}
\label{fig:chains}
\end{figure}

Definitely, the monopoles and the vortices are parts of a genuine non-Abelian object. For example, in $SU(2)$ gauge theory the center vortex
can be regarded as an Abelian vortex carrying the magnetic flux which is equal to a half of the total magnetic flux of a monopole.
The distribution of the magnetic part of the gluon energy density around a monopole is not spherical:
each monopole is a source of two vortex fluxes which must be connected to other anti-monopole(s)
because of a conservation of the vortex flux~\cite{ref:chains2}. As a result, there appears
a closed set of the vortex segments which connect alternating monopoles and antimonopoles, Figure~\ref{fig:chains}.
In SU($N_c$) gauge theories the monopoles and vortices form nets~\cite{ref:review.monopoles.vortices}.
Similar monopole-vortex chains were found in numerous (non-)supersymmetric non-Abelian gauge theories
involving various Higgs fields~\cite{ref:monvort}.

A general phase diagram of the monopole component is suggested to be as follows~\cite{chernodub}:
\begin{itemize}
\item $0 < T < T_c$: The monopoles form a condensate in the confinement phase.
\item $T = T_c$: The condensate melts into a monopole liquid at the phase transition.
\item $T_c < T \lesssim 2 T_c$: The liquid exists in the ``strongly coupled'' region in the deconfinement.
\item $T \approx 2 T_c$: The point of a gas--liquid crossover.
\item $T \gtrsim 2 T_c$:
As temperature increases the monopole liquid gradually evaporates into a gas.
\end{itemize}
The intermediate monopole liquid state was also discussed in Refs.~\cite{chernodub,shuryak} while the formation of the monopole gas
at very high temperatures was predicted in Refs.~\cite{chris,ref:blocking}.

\section{Thermodynamics of Yang-Mills theories and topological objects}

This Section is based on results of Ref.~\cite{nakamura}. Basic thermodynamical quantities of
Yang-Mills theory can be found from the expectation value of the trace $\theta$ of the energy--momentum tensor~$T_{\mu\nu}$:
\beqn
\theta(T) = \langle T^\mu_\mu \rangle \equiv \varepsilon - 3 p\,,
\qquad
T_{\mu\nu} = 2 \, \Tr \left[G_{\mu\sigma} G_{\nu\sigma} - \frac{1}{4} \delta_{\mu\nu} G_{\sigma\rho} G_{\sigma\rho}\right]\,,
\label{eq:anomaly:continuum}
\eeqn
where $G_{\mu\nu} = G_{\mu\nu}^a t^a$ is the field strength tensor of the gluon fields $A_\mu$. For example,
the pressure $p$, the energy density $\varepsilon$, and entropy $s$ can be calculated from the trace $\theta$ as follows:
\beqn
p(T) = T^4 \int\limits^T \ \frac{{\mathrm{d}}\, T_1}{T_1} \ \frac{\theta(T_1)}{T_1^4}\,,
\qquad
\varepsilon(T) = 3 \, p(T) + \theta(T)\,,
\qquad
s(T) = \frac{p(T) + \varepsilon(T)}{T}\,.
\label{eq:pressure:anomaly}
\eeqn

\begin{figure}[htb]
\begin{center}
\begin{tabular}{ll}
\includegraphics[scale=0.22,clip=false]{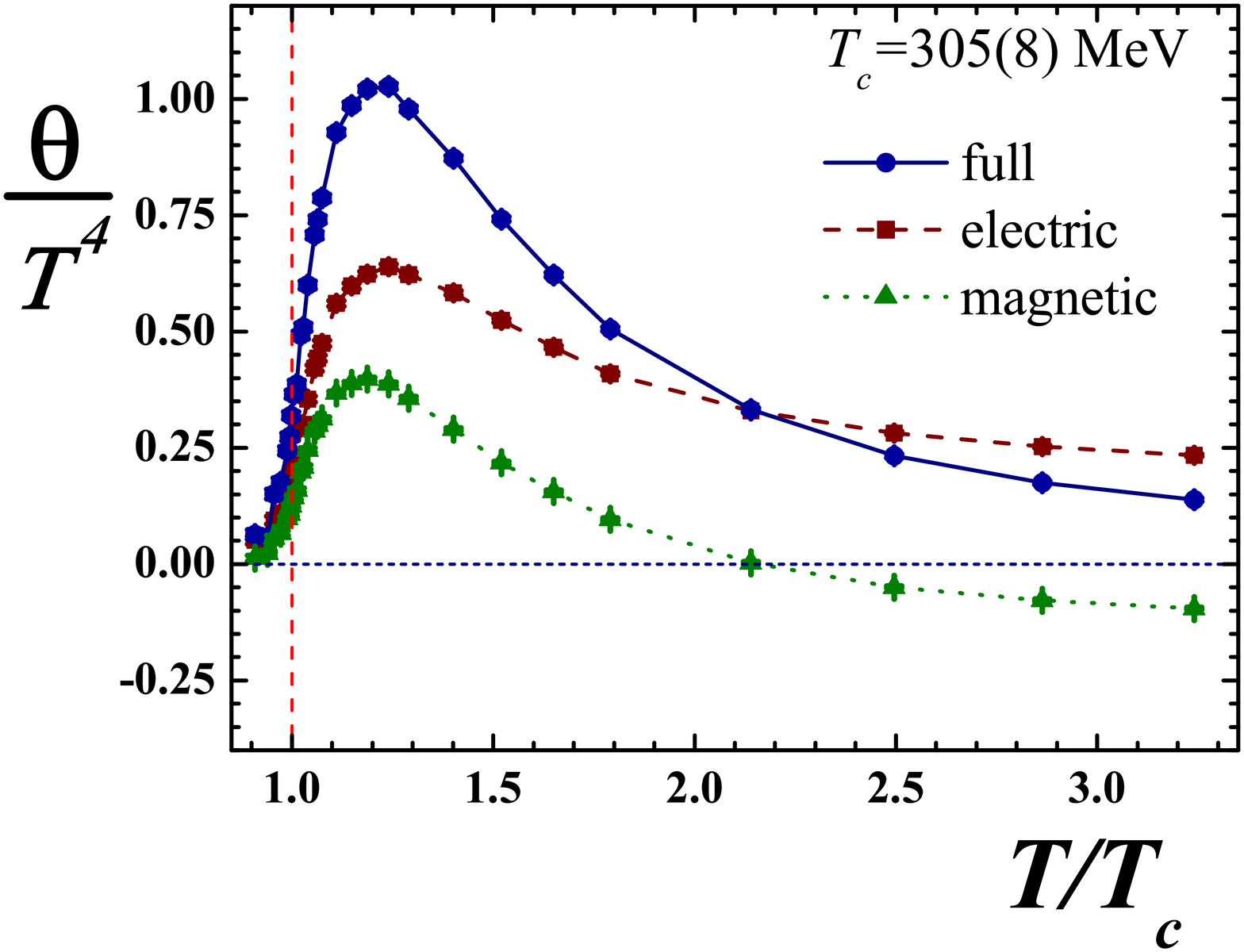}
& \\[-51mm]
& \includegraphics[scale=0.225,clip=true]{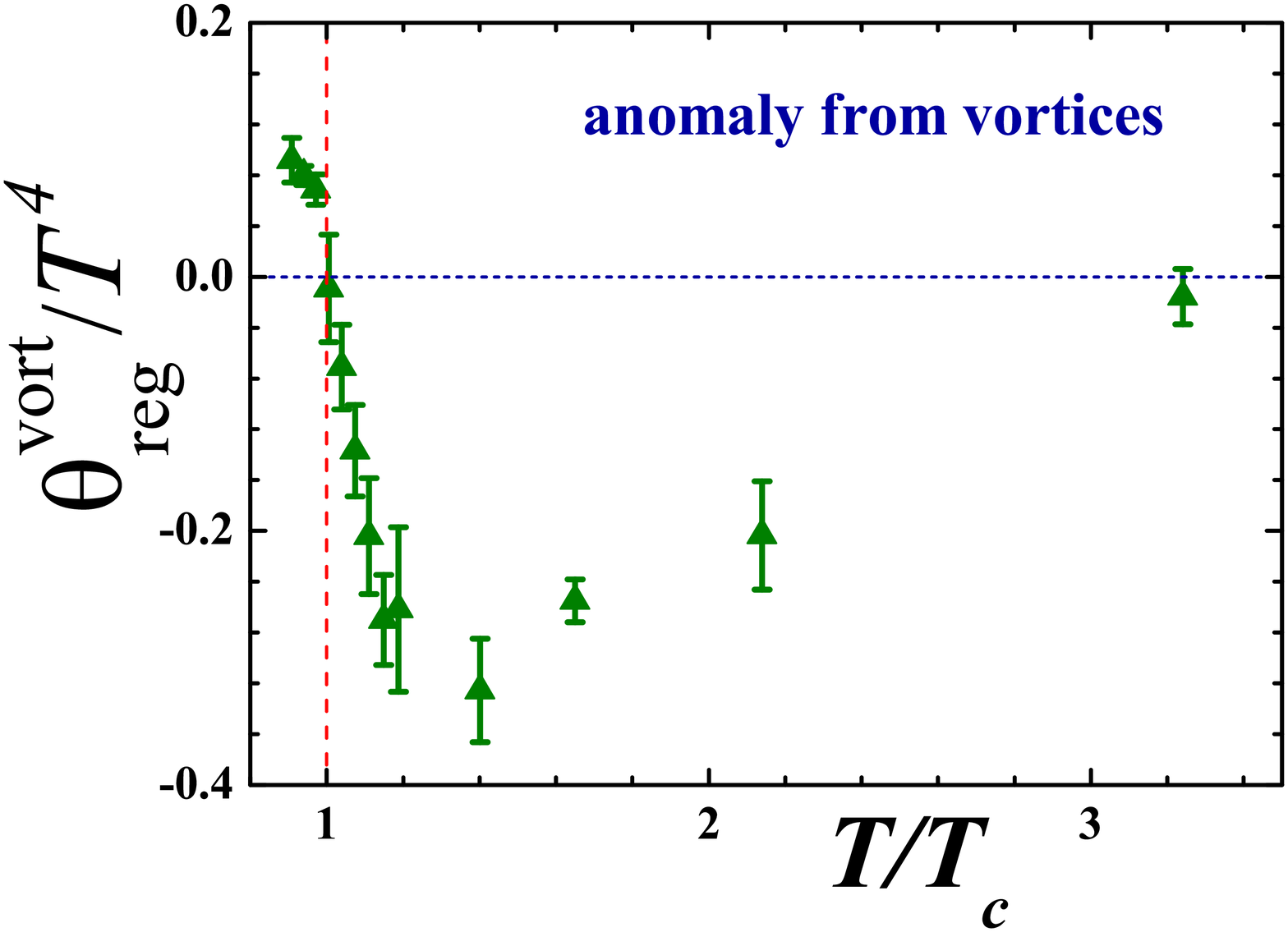}
\end{tabular}
\end{center}
\caption{(left) The trace anomaly $\theta$ (normalized by $T^4$)
as a function of the temperature~$T$ (in units of the critical temperature, $T_c$).
The full anomaly (circles), and its electric (squares) and magnetic (triangles) parts are shown.
(right) The contribution of the magnetic vortices into the trace anomaly.}
\label{fig:su2}
\end{figure}

The {\it bare} Yang--Mills theory is a conformal theory and therefore at the classical level the energy--momentum tensor
is traceless. However, because of a dimensional transmutation the energy--momentum tensor exhibits a trace anomaly,
\beqn
\theta = \left\langle \tilde\beta(g)\Tr G_{\mu\nu}^2(x) \right\rangle\,,
\qquad
\tilde\beta(g) \equiv \frac{\beta(g)}{g}  = \frac{{d} \log g}{{d} \log \mu} = - g^2 (b_0 + b_1 g^2 + \dots)\,.
\label{eq:theta}
\eeqn
In Figure~\ref{fig:su2} (left) we show the trace anomaly~\eq{eq:theta} for SU(2) gauge theory.
The measurements were performed for one lattice geometry, $18^3 \times 4$, and the scaling properties
of our results have not been studied yet.
The trace of the energy-momentum tensor
is subdivided into its electric (proportional to $G_{4i}^2$, $i=1,2,3$) and magnetic ($G_{ij}^2$, $i,j=1,2,3$) parts.
In Figure~\ref{fig:su2} (right) we show the contribution of the magnetic vortices into the trace anomaly of the gluon plasma.
In the deconfinement phase just above the phase transition (i.e., in the monopole liquid region) the
vortex-generated anomaly takes a negative value in agreement with general theoretical expectations~\cite{sasha}.

Thus, the vortex constituents of the monopole-vortex chains are relevant for the thermodynamics.
The monopole constituents are also important for thermodynamics~\cite{ref:review.monopoles.vortices,nakamura}
as they carry an excess of the (magnetic part of) non-Abelian action density~\cite{ref:physical}.
The action density, in turn, contributes to the trace anomaly~\eq{eq:theta} and, consequently, to the
pressure and to the energy density of the Yang-Mills plasma~\eq{eq:pressure:anomaly}.

\section{Effective 3d models of magnetic component}

In this Section we briefly compare the 4d, lattice-based  picture of the magnetic
component~\cite{chernodub,nakamura} with newly developed 3d models of the plasma.
We find a close relation between the two approaches, originally developed in
absolutely independent ways. The presentation is aimed to emphasize
rather the general picture emerging rather than details.

It is quite common nowadays to assume that the dynamics of the Polyakov's lines plays
$\Omega({\bf x})$ a crucial role in the confinement-deconfinement phase
transition~\cite{polyakov.yaffe,pisarski,kurkela}
\begin{equation}\label{omega}
\Omega({\bf x})~\equiv~P\exp\Big[-i \int_0^{1/T} \, d\tau A_0(\tau,{\bf x})\Big]
~~,
\end{equation}
where $\tau$ is the Euclidean time. The vacuum expectation value,
$\langle L \rangle$ with $L \equiv \frac{1}{2} \Tr \Omega$, serves as an order parameter,
which is vanishing in the confinement phase and is finite in the deconfining phase.
The symmetry -- which is violated by a nonvanishing vacuum expectation value $\langle L \rangle$
-- is the global $\Z_2$ symmetry (for definiteness we consider SU(2) gauge group).

Note that $\Omega({\bf x})$ depends only on 3d variables and we are invited
to consider 3d reduced, or effective theories.
A particular form of such a $\Z_2$--symmetric lagrangian looks as~\cite{pisarski} :
\begin{eqnarray}\label{reduced}
L^{\mathrm{eff}}(A_i , \Omega) = {1\over 2} \Tr G_{i,j}^2 +
{T^2\over g^2} \Tr|\Omega^{\dagger} D_i \Omega|^2 +V(|\Tr\Omega|^2),
\end{eqnarray}
where the potential $V(|\Tr \Omega|^2)$ can produce a non-trivial vacuum expectation
value, $\langle L \rangle \neq 0$.

The Lagrangian (\ref{reduced}) is nonrenormalizable in 3d. A renormalizable version of
the effective Lagrangian was suggested in Ref.~\cite{kurkela}. The
idea can be represented as follows.
There are two basic elements inherent to the construction
(\ref{reduced}). First, $\Z_2$ invariance of the Lagrangian and, second,
spontaneous breaking of the symmetry due to the potential.
Both elements can be realized in terms of local fields, rather than non-local
objects $\Omega$. To this end one introduces color triplet and color singlet
3d scalar fields $\Pi_a$ and $\Sigma$. The Higgs-fields Lagrangian is then the
standard kinetic terms plus the potential energy~\cite{kurkela}:
\begin{equation}\label{new}
V (\Sigma,\Pi_a )= b_1 \Sigma^2 + b_2 \Pi_a^2 + c_1\Sigma^4 + c_2 (\Pi_a^2)^2 + c_3 \Sigma^2 \Pi_a^2 \,,
\end{equation}
where the coefficients $b_{1,2}$ and $c_{1,2,3}$ are organized in such a way
that $\langle \Sigma \rangle \neq 0$.
The success of the 3d effective theory (\ref{new}) is impressive
both in terms of its numerical match to the original 4d theory~\cite{kurkela}
and in the clarity of the underlying symmetry-based argumentation for
its introduction.

We propose to identify the 3d color-singlet with the 3d projection
of the magnetic vortices (or, of the 4d magnetic component of the plasma)
onto the 3d space. We suggest that such an identification allows us
at least relate the newly made observations  to the known properties of the
4d magnetic component.

The vortices are 2d surfaces percolating at low temperatures.
The 3d projection of the vortices is given by intersections of the 2d surfaces with a 3d time slice.
This intersection is given, in general, by 1d defects, or lines. These lines, or trajectories,
are closed. The properties of the 1d defects were studied in detail~\cite{langfeld}.
It was shown that properties of the 2d and, consequently, of their 1d projection
depend crucially on the temperature in the vicinity of $T_c$. Namely,
the surfaces become time-oriented at $T>T_c$ and  do not percolate
any longer from the 4d point of view. However, as is emphasized in \cite{langfeld}
the 3d percolation continues, now in terms of the 1d defects.
Using the percolation theory one can readily argue, that the
lattice observations imply the inequality $\langle \phi_M\rangle  \neq 0$, 
where by $\phi_M \sim \Sigma$ we understand now a 3d field corresponding 
to the clusters of the mentioned 1d magnetic defects. A search for deeper 
links between these two approaches is underway.

\end{document}